\documentstyle[aps,prb,multicol]{revtex}
\input{epsf}

\def\cdag#1#2{c^{\dagger}_{#1 #2}}
\def\path#1#2#3{{\Phi(\qaxis{#1}\qaxis{#2}\qaxis{#3})}}
\def\c#1#2{c_{#1#2}}
\def\abs#1{ \mid #1 \mid}
\def\cross{\times}
\def\tendsto{\rightarrow}
\def\half{{\small \frac{1}{2}}}
\def\quarter{{\small \frac{1}{4}}}
\def\eighth{{\small \frac{1}{8}}}

\def\R#1{R_{#1}}
\def\qaxis#1{\widehat{\Omega}_{#1}}
\def\hc{{\rm h.c.}}
\def\utrans#1{U_{\qaxis{#1}}}

\begin{document}

\title{ Fictitious fluxes in doped antiferromagnets }

\author{ Stellan \"Ostlund and Martin Andersson }

\address{
Department of Theoretical Physics and Mechanics, \\
Chalmers University of Technology and G{\" o}teborg University, \\
S-412 96 G\"{o}teborg, Sweden}

\date{\today}

\maketitle
\widetext


\begin{abstract}
In a tight binding model of charged spin-$\half$ electrons on a square
lattice, a fully polarized ferromagnetic spin configuration generates
an apparent $U(1)$ flux given by $2\pi$ times the skyrmion charge
density of the ferromagnetic order parameter. We show here that for an
antiferromagnet, there are two ``fictitious'' magnetic fields, one staggered
and one unstaggered.  The staggered topological flux per unit cell can
be varied between $-\pi\le\Phi\le\pi$ with a negligible change in the
value of the effective nearest neighbor coupling constant whereas the
magnitude of the unstaggered flux is strongly coupled to the magnitude
of the second neighbor effective coupling. 
\end{abstract} 


\begin{multicols}{2}
\narrowtext
\section*{General properties of spin generated fluxes}
\label{sec:flux}

It is well known that in the theory of the quantum Hall problem
spin textures can generate an effective $ U(1) $ flux which 
acts as an effective magnetic field leading to 
an association between  topological and electrical charge. 
\cite{ref:hansson} 
In this paper we investigate to what extent this effect can  
be generalized to the antiferromagnet.

For definiteness, we consider the $t$-$J$ model on a square lattice. 
This model is described by
the Hamiltonian
\begin{equation}
H =  \; {\displaystyle \sum_{\langle ij\rangle} } \; 
	\left[ -(t^{ij} \cdag{i}{\sigma} \c{j}{\sigma}+\hc ) + J^{ij} 
	({\mathbf S}_{i}  \cdot {\mathbf S}_{j} - \quarter n_i n_j) \right]
\end{equation}
where the summation runs over nearest neighbor pairs $ \langle ij\rangle$ and 
$ {\mathbf S}_i =  \cdag{i}{\alpha} \vec{\sigma}_{\alpha \beta}\c{i}{\beta} $
and $\vec{\sigma}$ denotes the vector of Pauli matrices. All states containing 
doubly occupied sites have been excluded from the Hilbert space. [For 
a general reference, see Ref. 
\onlinecite{ref:fradkin}.] 

To make explicit the connection between spin rotations and effective
couplings, we introduce a local change of spin coordinates
[see for instance Ref. \onlinecite{schulz90}],
choosing the local spin quantization axis at site $ i $ along
$ \qaxis{i}$. The action of this local $SU(2)$ transformation on $ \c{i}{} $ 
is written
\begin{equation}
\c_{i}{} \longrightarrow \utrans{i} \c_{i}{} ,
\end{equation}
where 
\begin{equation}
\label{eq:udef}
\utrans{i}\sigma_z\utrans{i}^{\dagger}=\qaxis{i}\cdot\vec{\sigma}.
\end{equation}
The specification of $ \qaxis{i} $ fixes $\utrans{i} $ only up to an overall 
rotation about the new local $ z $-axis. Choosing 
$G_i=\exp [ -i\half\alpha_i \qaxis{i}\cdot\vec{\sigma} ] $ makes 
$\utrans{i}' = G_i \utrans{i}$
also satisfy the defining relation, Eq. (\ref{eq:udef}). To fix this remaining
degree of freedom, we arbitrarily choose $ \utrans{i} $ to correspond to a 
rotation about an axis lying in the spin $x$-$y$ plane. Defining the unit 
vector $ \widehat{\omega}_i =  \widehat{ ( \widehat{z} \cross {\qaxis{i}} ) } 
= (-\sin\phi_i,\cos\phi_i,0)$ we have
\begin{equation}
\begin{array}{ll} \label{eq:su2rot}
\utrans{i}  & = \exp [ i \half\theta_i \widehat{\omega}_i \cdot 
		\vec{\sigma} ] \\
   	    & = \cos{(\half\theta_i)} + i \sin{(\half\theta_i)}
		\widehat{\omega}_i \cdot \vec{\sigma} ,
\end{array}
\end{equation}
where $ \cos{\theta_i} = \widehat{z} \cdot \qaxis{i} $.

We then find the following Hamiltonian in the new spin coordinate system 
\begin{eqnarray} \label{eq:mess}
H &=& \sum_{\langle ij \rangle} \Bigl[ -
 ( t^{ij} \cdag{i}{\alpha} M^{ij}_{\alpha \beta } \c{j}{\beta}  + \hc ) \cr
 &&\hspace{7mm}+
J^{ij} ( S^{\alpha}_i S^{\beta}_j Q^{ij}_{\alpha \beta } - \quarter n_i n_j )
   \Bigr] ,
\end{eqnarray}
where we have introduced $ M^{ij} = (\utrans{i})^{\dagger} \utrans{j} $, 
$ Q^{ij} = \R{\qaxis{i}}^{-1} \R{ \qaxis{j} } $, and
$(\R{\qaxis{}})_{ij} = \cos\theta  \delta_{ij}  +  (1-\cos{\theta}) \omega_i 
\omega_j + \sum_k\sin{\theta} \epsilon_{ijk} 
\omega^k $ is the $SO(3)$-rotation operator induced by $\utrans{}$.

Until this point, the discussion has been completely general. In order 
to make further progress, we neglect spin fluctuations and
make the restriction that a site is occupied by at most one spin and
that the electron which occupies site $ i $ has its spin pointing
along the local positive $z$-axis, $\widehat{\Omega}_i$. This contraint 
reduces Eq. (\ref{eq:mess}) to 
\begin{equation} \label{eq:electrongas}
H_{\rm eff} = {\displaystyle \sum_{\langle ij \rangle} } \; \left[
	- ( {\tau}^{ij}  \cdag{i}{} \c{j}{} + \hc ) +
	  K^{ij} n_i n_j \right] ,
\end{equation}
with $ {\tau}^{ij}  = t^{ij} M^{ij}_{11} $,
$ K^{ij} = \quarter J^{ij} (\qaxis{i} \cdot \qaxis{j} - 1)$, and
$ \c{j}{} \equiv \c{j}{1} $. At this point we have an effective model
describing ``spinless'' fermions on a lattice with hopping amplitudes and
interaction-strengths being functions of position. In this way we have
automatically solved the constraint of no double occupancies at the 
expense of treating the spins as classical variables and ignoring their 
quantum spin fluctuations.

Let us now turn to the properties of $ {\tau}^{ij} $. First we note that 
$ {\tau}^{ij} = ( {\tau}^{ji})^* $. The complex phase of $ {\tau}^{ij} $ 
cannot in general be gauged away by a local transformation 
$ \c{j}{}\longrightarrow e^{i \phi_j } \c{j}{} $ if the spin configurations 
are noncoplanar.  When we defined the local transformation $\utrans{i}$, 
we noted that it was
only specified up to a rotation about the local $z$-axis, 
$\utrans{ }\longrightarrow G\utrans{ }$.
The effect of such a local rotation on the effective hopping cannot be 
distinguished from a local gauge transformation 
$ \c_i\longrightarrow e^{i \half \alpha_i}\c_i$ which does not affect
the physics. Hence, the set of physically inequivalent choices of $\utrans{i}$
belong to $SU(2)/U(1)\cong S^2$ and we conclude that in the absence of
an external magnetic field, there are two physical 
degrees of freedom per site (or plaquette) which determine the effective 
coupling constants. 

To understand more precisely which portions of the effective hopping 
$ {\tau}^{ij} $ are $ U(1) $ gauge invariant,  we look at a particular 
plaquette of the lattice, consisting of points labelled counterclockwise as 
$ r_0, r_1, r_2, $ and  $ r_3 $ and with associated spins pointing in the
$ \Omega_{i} $ directions.  The flux through the plaquette is given by
\begin{equation}
\Phi_{plaquette} = {\rm Im}\ln \left( {\tau}^{r_0 r_1} {\tau}^{r_1 r_2} 
			{\tau}^{r_2 r_3} {\tau}^{r_3 r_0 }  \right) .
\label{eq:fluxplaq}
\end{equation}
Topological arguments show that
$ \Phi_{plaquette} = \half { \mathcal A}( \qaxis{0},\qaxis{1},\qaxis{2},
\qaxis{3} ) $, half the solid angle enclosed by the shortest path on the 
sphere connecting the vectors $\{ \qaxis{i} \}$, i.e. the flux corresponding 
to the plaquette is equal to $2\pi$ times the skyrmion charge represented by 
the plaquette. 
Generalizing a result of Wen, Wilczek, and Zee
\cite{wilczek}
who computed the imaginary part of the formula below
one may show that for a plaquette consisting of exactly three sites
\begin{eqnarray}
{\tau}^{r_0 r_1} {\tau}^{r_1 r_2} {\tau}^{r_2 r_0} 
	&=& \eighth \Bigl[(\qaxis{0}+ \qaxis{1}+ \qaxis{2})^2-1\Bigr] \cr 
	&&+ \frac{i}{4} \qaxis{0} \cdot ( \qaxis{1} \times \qaxis{2} ) .
\end{eqnarray}

Assuming smooth fields $\theta ({\mathbf r})$ and $\phi ({\mathbf r})$ one
finds in the continuum limit that the fictitious flux corresponds to a 
Berry gauge field 
$$
{\cal A}_{\mu} 	= \langle \widehat{\Omega} |
		  \partial_{\mu} | \widehat{\Omega} \rangle 
		= i \sin^2 \frac{\theta}{2} \partial_{\mu}\phi ,
$$
which is the vector potential due to a magnetic monopole of strength
$-\half$. The local $U(1)$ degree of freedom, represented by the $G$'s, 
corresponds to a gauge transformation of the topological vector potential 
${\cal A}_{\mu}$.

 
\subsection*{The case of the ferromagnet}

\begin{figure}[tbh]
\centerline{\epsfxsize=0.6\columnwidth\epsffile{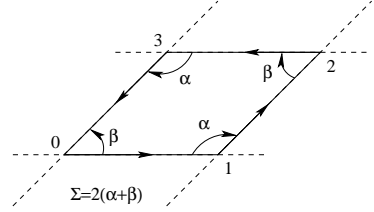}}
\vspace{0.5cm}
\caption[ferro]{\label{fig:ferro}
The close to ferromagnetic spin-configuration is shown in
spin-space. All sides in this spherical parallelogram correspond to an 
opening angle $\theta$. $\Sigma$ denotes the sum of the angles within the
path $ 0 \rightarrow 1 \rightarrow 2 \rightarrow 3 \rightarrow 0 $.
}
\end{figure}

Let us now consider the plaquette $ ( \qaxis{0},\qaxis{1},
\qaxis{2},\qaxis{3}) $ in spin space, drawn as in  Fig.~\ref{fig:ferro},
the parallelogram representing a patch of the surface of the sphere.
To simplify the argument further, we restrict ourselves
to the case $ \qaxis{i} \cdot \qaxis{j} = \cos{ \theta } $ for all nearest 
neighbor pairs $ \langle ij \rangle $ in the plaquette, so that the relative 
angle between the spins on each side of the square is $ \theta $.
A straightforward application of spherical geometry yields the relation 
$ \half \; {\mathcal A} =   ( \alpha + \beta ) - \pi  $ where $ \alpha $ 
and $ \beta $ are the interior surface angles of the spherical parallelogram. 
However,
since $ \theta $ is fixed on all sides, we cannot freely choose $ \alpha $ and 
$ \beta $. Rather, we solve for $\Phi =\half { \mathcal A} $ in terms of 
$ \alpha - \beta $ and $ \theta $ and then derive expressions for the
nearest-neighbor coupling constants $ \tau $ and $ K $ in terms of the same
variables. Using standard formulas from spherical geometry 
together with Eq. (\ref{eq:su2rot}) we find 
\begin{equation}
\label{eq:fmcoupling}
\begin{array}{ll}
\abs{\tau_{nn}} & =  \cos ( \half \theta ) \\
\Phi    	& =  2 \arcsin \left( \tan^2 ( \half \theta ) 
			\cos( \half ( \alpha - \beta ) ) \right) \\
K		& =  \quarter J ( \cos \theta - 1 ) ,
\end{array}
\end{equation} 
where $ {\tau}_{nn} $ is the effective nearest-neighbor hopping amplitude.
We note that $ \Phi $ is given by 
$ \Phi = \half \theta^2 \cos ( \half ( \alpha - \beta ) ) $
for small $ \theta $.


\subsection*{The case of antiferromagnetic order}

\begin{figure}[tbh]
\centerline{\epsfxsize=\columnwidth\epsffile{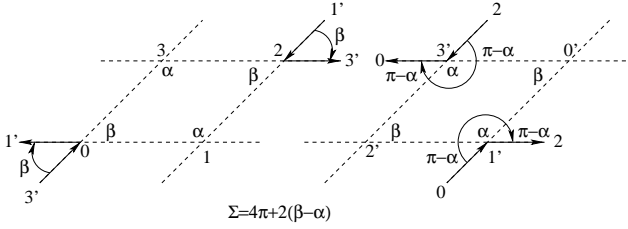}}
\vspace{0.5cm}
\caption[antiferro]{\label{fig:antiferro}
The close to antiferromagnetic spin-configuration is shown in 
spin-space. The left part corresponds to the points $\qaxis{i}$, while the
right part corresponds to the antipodal points $\qaxis{i}'$. It is seen how 
the $\alpha$-angles become exterior rather than interior leading to a 
completely different $\Sigma$ as compared to the ferromagnetic
case.
}
\end{figure}

\begin{figure}[tbh]
\centerline{\epsfxsize=0.6\columnwidth
\epsffile{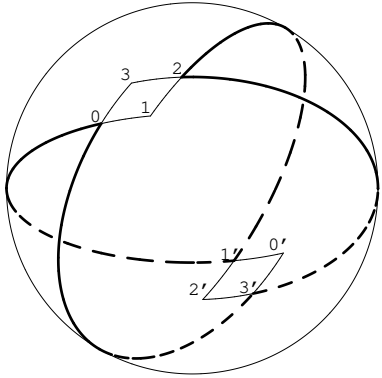}}
\vspace{0.5cm}
\caption[sphere]{\label{fig:sphere} 
The path taken in spin-space when going around a plaquette $0\tendsto 
1'\tendsto 2\tendsto 3'\tendsto 0$ in an antiferromagnetic background. All 
paths follow great circles on the sphere.
}
\end{figure}

Let us now consider the analogous calculations for a spin
configuration which is close to antiferromagnetic.  Let us denote
the local antiferromagnetic spin as $ \qaxis{i} $ as before. Denote
by $ \qaxis{i}^{\prime} = - \qaxis{i} $ the antipodal points.
Noting that in the case of a spherical parallelogram, great circles
which connect the four sides of $  \{ \qaxis{i} \} $  intersect
in the points  $ \{ \qaxis{i}^{\prime} \} $, we find that the
path $   \qaxis{0} \rightarrow  \qaxis{1}^{\prime} \rightarrow \qaxis{2} 
\rightarrow \qaxis{3}^{\prime} \rightarrow \qaxis{0} $ 
connecting the antiferromagnetic spins has the geometry sketched in
Fig.~\ref{fig:antiferro}. Fig.~\ref{fig:sphere} shows another illustration of 
the path taken in spin space when going around a plaquette in an 
antiferromagnetic background. By using Eq. (~\ref{eq:su2rot}) the angles 
$ \alpha $ have become exterior rather than interior angles in the path. This 
gives us the relations

\begin{equation}
\begin{array}{ll}
\abs{\tau_{nn} } 	& =	\sin ( \half  \theta) \\
\Phi    		& =  	\pi - ( \alpha - \beta ) \\
K       		& =  	- \quarter J (\cos \theta + 1 )
\end{array}
\end{equation}
where in this case $ \cos \theta = - \qaxis{i} \cdot \qaxis{i+1}' $. Comparing
to Eq. (~\ref{eq:fmcoupling}) we see that the connection between $\theta$ and
$\Phi$ has disappeared.
By extending the same argument to the adjacent cell, we reproduce the 
arguments, however all paths are traversed in the opposite sense and the flux 
becomes negative compared to the first cell . The flux is therefore staggered,
with a $ \pm $ sign depending on the sign of the sublattice associated with 
the plaquette.

In contrast to the case of the ferromagnet, we find that we can 
change the effective staggered flux for the
electron gas, Eq. (\ref{eq:electrongas}), from $ -\pi $ to $ \pi $ without 
affecting the magnitude of the 
local values of the effective nearest neighbor coupling constants. In fact, 
for the most symmetric case $ \alpha = \beta $, we find that an 
antiferromagnetic skyrmion generates an effective staggered flux of exactly 
$ \pi $ per plaquette. Due to the curvature of the sphere, such a texture 
cannot in general
be extended to the plane.


\subsection*{Next-nearest neighbor hopping}

When  we include second neighbor hopping, the situation becomes  much more 
complicated for the antiferromagnet. Each square plaquette has four 
gauge invariant fluxes, 
corresponding to each triangle defined by the removal of one vertex from the 
four corners of the square. When the spins are aligned close to
ferromagnetically, the shortest path in spin space connecting the two diagonal
spins lies wholly within the region defined by the corner spins. Hence the 
flux through each triangular subplaquette is very closely proportional to the 
area of each subtriangle in real space, and the sum of flux through a pair of 
subtriangles that cover the square is precisely equal to the flux through
the entire plaquette. As a consequence, the effective phase of each of the 
interactions is extremely well approximated by spreading a constant effective 
magnetic field corresponding to the local skyrmion density throughout the 
entire real space plaquette and assigning the effective phase on the links by 
a conventional choice of gauge\cite{barford91}.

In the case of a configuration close to antiferromagnetic, the four 
subtriangles cover the entire sphere in spin space [see 
Fig.~\ref{fig:sphere}]. Taking into account the 
orientation of each of the bounding paths, we find the following relation 
between the flux through each of the triangles for both the ferromagnet and 
antiferromagnet:
\begin{eqnarray}  
\label{eq:fluxc}
 \path{0}{1'}{2} &+& \path{0}{2}{3'} \cr 
	&-& \path{0}{1'}{3'} - \path{1'}{2}{3'} 
	= 2 \pi  n ,
\end{eqnarray}
where $ n = 0 $ for the ferromagnet and $ n = \pm 1 $ for the antiferromagnet.
Note that Eq. (~\ref{eq:fluxc}) is valid in general, i.e. it does
not rely on the assumption of a spherical parallelogram. It is also easy to
see that it is valid in the presence of an external electromagnetic flux.
If we assume a 
spherical parallelogram we can also note from Fig.  ~\ref{fig:antiferro} that 
$ \path{0}{1'}{3'}  = \path{1'}{2}{3'} = \half \Phi  $. It therefore follows 
that $ \path{1'}{2}{0}  = \path{0}{2}{3'} = \half \Phi + n  \pi $. Hence, 
for the antiferromagnet, links connecting sites within the same sublattice
pick up a phase $\pi$ if they belong to sublattice $A$, and zero if they belong
to sublattice $B$, or vice versa.  
\cite{ref:regauge}

\begin{figure}[tbh]
\centerline{\epsfxsize=0.7\columnwidth
\epsffile{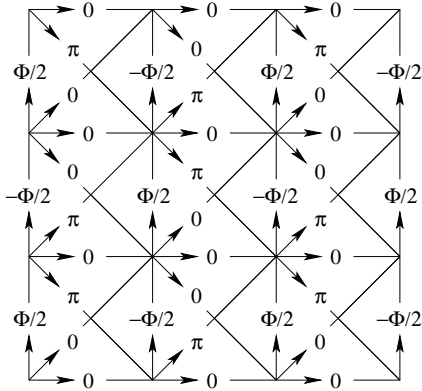}}
\vspace{0.5cm}
\caption[lattice]{\label{fig:lattice} 
A particular choice of gauge for the staggered flux-phase ($\pm\Phi$ )
with next-nearest neighbor hopping is indicated through the phases associated 
with each link in the lattice. 

}
\end{figure}

In Fig. ~\ref{fig:lattice} is drawn a 
$3\times 3$ subset of a 2D antiferromagnetic lattice. The value of the local
spin generated phase $ {\rm Im} \ln ( M^{ij}_{11} ) \equiv  \phi_{ij} $ is 
indicated on each link. Arrows indicate the direction $ i \rightarrow j $. The
gauge is chosen so that $ \phi_{r,r+\widehat{x}}  = 0 $. The phase 
$ \phi_{r,r+\widehat{y}} =  \frac{(-1)^r}{2} \Phi $. Along the diagonals, 
the phase of the coupling constant is:
\begin{equation}
\begin{array}{ll}
\phi_{r , r + \widehat{y} + \widehat{x} }   	& = \frac{\pi}{2}[1+(-1)^r] \\
\phi_{r + \widehat{x} , r + \widehat{y} }   	& = \frac{\pi}{2}[1-(-1)^r] .\\
\end{array}
\end{equation}

The local gauge choice does not take account of the curvature of the spin 
space and in general, the sphere cannot be covered with parallelograms so 
the local choice of gauge cannot be extended to the whole sphere. 
Nevertheless, 
in the limit $ \theta \tendsto 0 $ the above  formula is expected to be valid.

An intriguing special choice for $\theta$ where it is
possible to cover the sphere by parallelograms is $\cos\theta=-1/3$. In this
case, the spins reside on the vertices of a regular tetrahedron, and it is
also possible to extend this mapping exactly over a square lattice. 
Considering that 
going around a plaquette in the real space lattice, corresponds to encircling
two of the faces of the tetrahedron, it is obvious that this loop will cover
half of the spin-sphere and hence the ``tetrahedral phase'' generates a 
staggered flux $\pm\pi$.

Putting all this together, we find that given original nearest ($t_{nn}$) and
next nearest neighbor hopping ($t_{nnn}$), the effective 
coupling constants for the remaining fermionic degrees of freedom are
determined by the two free parameters $ \theta $ and $ \Phi $ which control the
unstaggered and staggered fluxes, respectively. 
In the general case with an external electromagnetic flux we have $2+1$ 
physical degrees
of freedom per plaquette. This results in three independent fluxes per square,
the only restriction being given by Eq. (\ref{eq:fluxc}).
Using standard 
formulas from spherical geometry, we find that with a suitable choice of 
gauge, the effective nearest and next nearest neighbor couplings are given by:
\begin{equation} \label{nnn}
\begin{array}{ll}
\tau_{nn} 		& = t_{nn} \sin( \half \theta ) \\
\tau_{r + \hat{x} } 	& = \tau_{nn} \\
\tau_{r + \hat{y} } 	& = \tau_{nn} e^{ i (-1)^r \Phi / 2  } \\ 
\cos( \theta_{+} \pm \theta_{-}  ) 
			& = \cos( \theta ) \pm  ( 1 -  \cos( \theta ) ) 
			    \sin( \half \Phi ) \\
\tau_{r,r+\hat{x}+\hat{y} } 
			& =  (-1)^{r+1} t_{nnn} \cos(  \theta_{+} ) \\
\tau_{r+ \hat{y},r+\hat{x} } 
			& =  (-1)^r t_{nnn} \cos( \theta_{-} ). \\
\end{array}
\end{equation}

\section*{Conclusion}

The main conclusion from this paper is that in contrast to 
the ferromagnet, antiferromagnetic spin textures naturally generate
a staggered flux\cite{affleck87} 
on the basic nearest neighbor plaquettes.
However, in contrast to case of the ferromagnet, the amount of 
flux per plaquette is weakly related
to the {\it magnitude } of the effective nearest neighbor
coupling. Incorporating second neighbor interactions reveals the
existence of an additional independent effective $ U(1) $ fictitious
flux on plaquettes of second neighbor sites. Since 
the first neighbor coupling is severely suppressed in a N\'{e}el 
antiferromagnet, 
it is quite possible that the effective second neighbor interactions could 
be important; in the limit of weakly coupled effective nearest 
neighbors and strongly coupled second neighbors textures such as appear in the
two layer quantum Hall problem could be favored.\cite{girvin99}
However, even in this case the magnetic field energy generated by 
staggered orbital currents could suppress these textures.\cite{lubensky} 

Using Eq. (\ref{nnn}) we \cite{ref:andersson}
have numerically calculated energies of various
uniform spin textures in the  Hartree-Fock approximation and for
nonuniform stripe textures in the Hartree approximation, with the
goal of understanding if noncoplanar spin textures which
generate fictitious staggered flux have lower energy than the coplanar
``spiral'' phases studied previously by a number of authors.
\cite{schulz90,siggia92,eriksson}
Preliminary results are consistent with those that have been 
obtained previously by more elaborate means; spiral phases appear
to have the lowest energy for the uniform phases but are
thermodynamically unstable against phase separation. 
The noncoplanar phases all have staggered fictitious
rather than the unstaggered flux which is known to lower
the kinetic energy in an electron gas and there does 
not appear to be any clear association between topological charge and doping, 
at least in a tight binding model dominated by nearest neighbor
interactions. \cite{hasegawa89}

By neglecting quantum spin fluctuations as we have done, our
numerical  simulations  of the nonuniform spin textures suffer from 
the same problem as others who similarly ignore spin fluctuations. 
The ordinary N\'{e}el antiferromagnet becomes insulating, and  this, 
together with the tendency towards phase separation apparently does 
not favor noncoplanar spin textures with any nontrivial topological 
properties. It  remains to be seen if properly accounting for 
quantum spin fluctuations can change this result.


\end{multicols}

\begin{references}

\bibitem{ref:hansson}
S. C. Zhang, T. H. Hansson, and S. A. Kivelson, Phys. Rev. Lett. {\bf 62}, 82 
(1989);
A. Karlhede, S. A. Kivelson, K. Lejnell, and S. L. Sondhi, Phys. Rev. Lett. 
{\bf 77},
2061 (1996); D. H. Lee and C. L. Kane, Phys. Rev. Lett. {\bf 64}, 1313 (1990).

\bibitem{ref:fradkin}
See for instance E. Fradkin, {\it Field Theories of Condensed Matter Physics},
Addison Wesley (1991) and references therein.

\bibitem{schulz90}
H. J. Schulz, Phys. Rev. Lett. {\bf 65}, 2462 (1990).

\bibitem{wilczek}
The imaginary part of this expression was pointed out by 
X. G. Wen, F. Wilczek, and A. Zee, Phys. Rev. B. {\bf 39}, 11413 (1989).

\bibitem{barford91}
W. Barford and J. H. Kim, Phys. Rev. B {\bf 43}, 559 (1991).

\bibitem{ref:regauge}
By multiplying   $\cdag{r}{} $ by
$ (-1) $ if $ r \cdot \widehat{x} $ and $  r \cdot \widehat{y} $ are odd,
we can gauge away the staggered minus sign in the diagonal phases at the
price of introducing an additional staggered component in the
nearest neighbor couplings.

\bibitem{affleck87}
I. Affleck and J. B. Marston, Phys. Rev. B {\bf 37}, R3774 (1987).

\bibitem{girvin99}
S. M. Girvin, {\it The Quantum Hall Effect, Novel Excitations
and Broken Symmetries }, Les Houches Lectures, 1998.
Published by Springer Verlag (1999). See also
cond-mat/9907002.

\bibitem{lubensky}
A. B. Harris, T. C. Lubensky, and E. J. Mele,
Phys. Rev. B {\bf 40 }, 2631 (1989).

\bibitem{ref:andersson}
M. Andersson and S. \"Ostlund, (unpublished).

\bibitem{siggia92}
B. I. Shraiman and E. D. Siggia, Phys. Rev. B {\bf 46}, 8305 (1992);
{\it ibid } Phys. Rev. Lett. {\bf 61}, 467 (1991).

\bibitem{eriksson}
A. B. Eriksson, T. Einarsson, and S. {\" O}stlund, Phys. Rev. B {\bf 52}, 3662 
(1995).

\bibitem{hasegawa89}
Y. Hasegawa, P. Lederer, T. M. Rice, and P. B. Wiegmann, Phys. Rev. Lett. 
{\bf 63},  907 (1989).

\end{references}
\end{document}